\documentstyle[aps,preprint]{revtex}
\input psfig
\def\k{\mbox{\bf k}}
\def\be{\begin{equation}}
\def\ee{\end{equation}}
\def\half{\frac{1}{2}}
\def\ltap{\raisebox{-.55ex}{\rlap{$\sim$}} \raisebox{.4ex}{$<$}}
\def\gtap{\raisebox{-.55ex}{\rlap{$\sim$}} \raisebox{.4ex}{$>$}}
\def\gsim{\mathrel{\gtap}}
\def\lsim{\mathrel{\ltap}}
\begin{document}
\baselineskip.5cm
\parskip4pt
\draft
\preprint{PURD-TH-96-06, OSU-TA-20/96, hep-ph/9608458}
\date{August 27, 1996}
\title{The Universe after inflation: the wide resonance case}

\author{S. Yu. Khlebnikov$^1$ and I. I. Tkachev$^{2,3}$}
\address{
${}^1$Department of Physics, Purdue University, West Lafayette, IN 47907 \\
${}^2$Department of Physics, The Ohio State University, Columbus, OH 43210\\ 
${}^3$Institute for Nuclear Research of the Academy of Sciences of Russia
\\Moscow 117312, Russia}

\maketitle

\begin{abstract}
\baselineskip0.5cm
We study numerically the decay of massive and massless inflatons into 
massive excitations, via a $\phi^2 X^2$ coupling, in the expanding Universe.
We find that a wide enough resonance can survive the Universe
expansion, though account for the expansion is very important for determining 
precisely how wide it should be.
For a massive inflaton, the effective production of particles with mass ten 
times that of the inflaton requires very large values of the resonance
parameter $q$, $q\gsim 10^8$. For these large $q$, the maximal size of produced 
fluctuations is significantly suppressed by the back reaction, but 
at least within the
Hartree approximation they are still not negligible. 
For the massless inflaton with a $\lambda\phi^4/4$
potential, the Universe expansion completely prevents a resonance production 
of particles with masses larger than $\sqrt{\lambda}\phi(0)$ 
for $q$ up to $q=10^6$.
\end{abstract}


%\newpage
\narrowtext
The quantitative theory of reheating after inflation
is an important problem in inflationary cosmology \cite{infl}
because the way the reheating proceeds determines possibility of certain 
phase transitions and scenarios of baryogenesis. 
It was recently realized that stimulated processes, due to large
occupation numbers of produced bosons, can be most significant during 
reheating \cite{KLS,STB,Betal,us}.
However, despite considerable progress made \cite{KLS,F&al,Yoshimura},
still little was known quantitatively about the important case
when the coupling of the inflaton field ($\phi$) to bosonic fields ($X$)
is relatively large (the wide resonance case).
It was not clear precisely how wide the resonance should be (for a given mass
of $X$) to overcome the expansion of the Universe. If the coupling required by 
this condition turns out to be too large, the resonance will 
be prevented by non-linear effects, which in this case will set in when $X$ 
fluctuations are still of the quantum strength. 
Thus even the existence of a wide resonance in the expanding Universe was
not firmly established, let alone its quantitative details.
The purpose of the present work was to study the wide resonance case by 
numerical means. 

The main quantity that one wants to know for a given inflationary
model is the maximal size of fluctuations of fields produced in 
the inflaton decay. 
The size of these fluctuations determines which phase transitions took 
place after inflation \cite{KLS2,Tkachev}, 
whether enough of heavy leptoquarks could
be produced to generate the baryon asymmetry \cite{KLS2,Tkachev,Yoshimura,KLR}, 
and how strong was the cosmological breakdown of supersymmetry \cite{ALR,DR}.

The reason why fluctuations produced at reheating could be 
sufficiently large for the above effects to exist is that, as now recognized,
reheating in many cases starts with a stage of parametric resonance (called
preheating in ref. \cite{KLS}), during which the bosonic fluctuations
grow extremely (exponentially) fast. In the expanding Universe, 
a parametric resonance is effective only when it is sufficiently wide or when
the system is (classically) conformally invariant.
Otherwise, the redshift of momenta prevents the resonance from developing
\cite{axion}.
The conformally invariant case is realized when a massless inflaton decays
into a massless fluctuation field. Production of a massive field
can only take place in the wide resonance regime. In that case, there can occur
an interesting and somewhat counterintuitive 
phenomenon, when quanta of a field that strongly couples to the inflaton can be
efficiently produced even if their mass is much larger than the frequency of
the inflaton oscillations. 
The effect can be thought of as a ``many$\to$ few" process, 
in which many quanta
of the inflaton field assemble together to produce a few heavy quanta.
If the fields that can be efficiently produced by this mechanism include 
leptoquarks of a grand unified theory (GUT), the subsequent non-equilibrium 
decay of these leptoquarks could give rise to a sizable baryon asymmetry 
\cite{KLS2,Tkachev,Yoshimura,KLR}.

In the present work, we wanted, first of all,
to find the regions of parameters,
for which a wide resonance can overcome the redshift of fields and momenta
in the expanding Universe.
We considered two models: in the first one, the inflaton potential was 
$m^2\phi^2/2$, in the second, it was $\lambda\phi^4/4$. In both cases,
we considered decay of the inflaton into a heavy scalar
field $X$ of mass $M_X$, due to coupling $\half g^2 \phi^2 X^2$. 
The width of the resonance in these models is determined by parameter $q$:
in Model 1 (a massive inflaton) $q\equiv g^2 \phi^2(0)/4 m^2$, where 
$\phi(0)$ is the value of the inflaton field when it starts to 
oscillate;\footnote{
For a static universe, in the linear regime with respect
to fluctuations, this agrees with the definition of
the parameter $q$ of the canonical Mathieu equation \cite{book}.} 
in Model 2 (the massless inflaton), $q\equiv g^2/4\lambda$. 

In Model 2, we have found no resonance for any $M_X > \sqrt{\lambda}
\phi(0)$ with
$q$ up to $q=10^6$. In Model 1 with $M_X/m=10$, which may be
appropriate for scalar leptoquarks, we have found that the redshift
is overcome only for very large $q$, $q\gsim 10^8$.
Thus, a wide enough resonance can indeed survive in the expanding Universe, 
though account for the expansion is very important for determining precisely 
how wide it should be.

Our second objective in this work was to determine the maximal size achieved
by $X$ fluctuations in Model 1. 
In general, both the Universe expansion and the back reaction play
a role in determining the maximal size of fluctuations.
A fully non-linear calculation along the lines of ref. \cite{us}, which 
would allow us to include consistently all back reaction effects, was
technically unfeasible for $q\gg 10^3$. So, we took the back 
reaction into account approximately, 
using a simple Hartree procedure \cite{KLS,Betal}.

The Hartree approximation may overestimate the maximal size of fluctuations.
The interaction term in the equation of motion for $\phi$ becomes important
when $\langle X^2 \rangle\sim \phi^2(0)/q$ (the brackets denote averaging 
over space). One might expect that this value signifies the end of the 
resonance stage. However, in the Hartree approximation,
when this value is reached, $\langle X^2 \rangle$ merely renormalizes the
mass of $\phi$ \cite{KLS} and continues to grow. We cannot say at present
if that further growth is an artifact of the Hartree approximation in this 
model and will disappear in the fully non-linear picture. In the conformally 
invariant case (see below), 
the limited set of data points that we have for the full problem does 
fit on a $1/q$ curve, in disagreement with the Hartree approximation.
We are not willing to make any
inferences for the present case, however, because the time evolution of
is rather model dependent.

We neglected any self-coupling of the field $X$. In the Hartree approximation,
an estimate shows that a quartic self-coupling ${\tilde \lambda} X^4/4$
can be neglected during the entire resonance stage when
${\tilde \lambda}\lsim g^2$. If the maximal size of fluctuations is in fact
smaller than obtained in the Hartree approximation, this self-coupling can be
neglected at the resonance stage even for larger ${\tilde \lambda}$.

\begin{figure}
\psfig{file=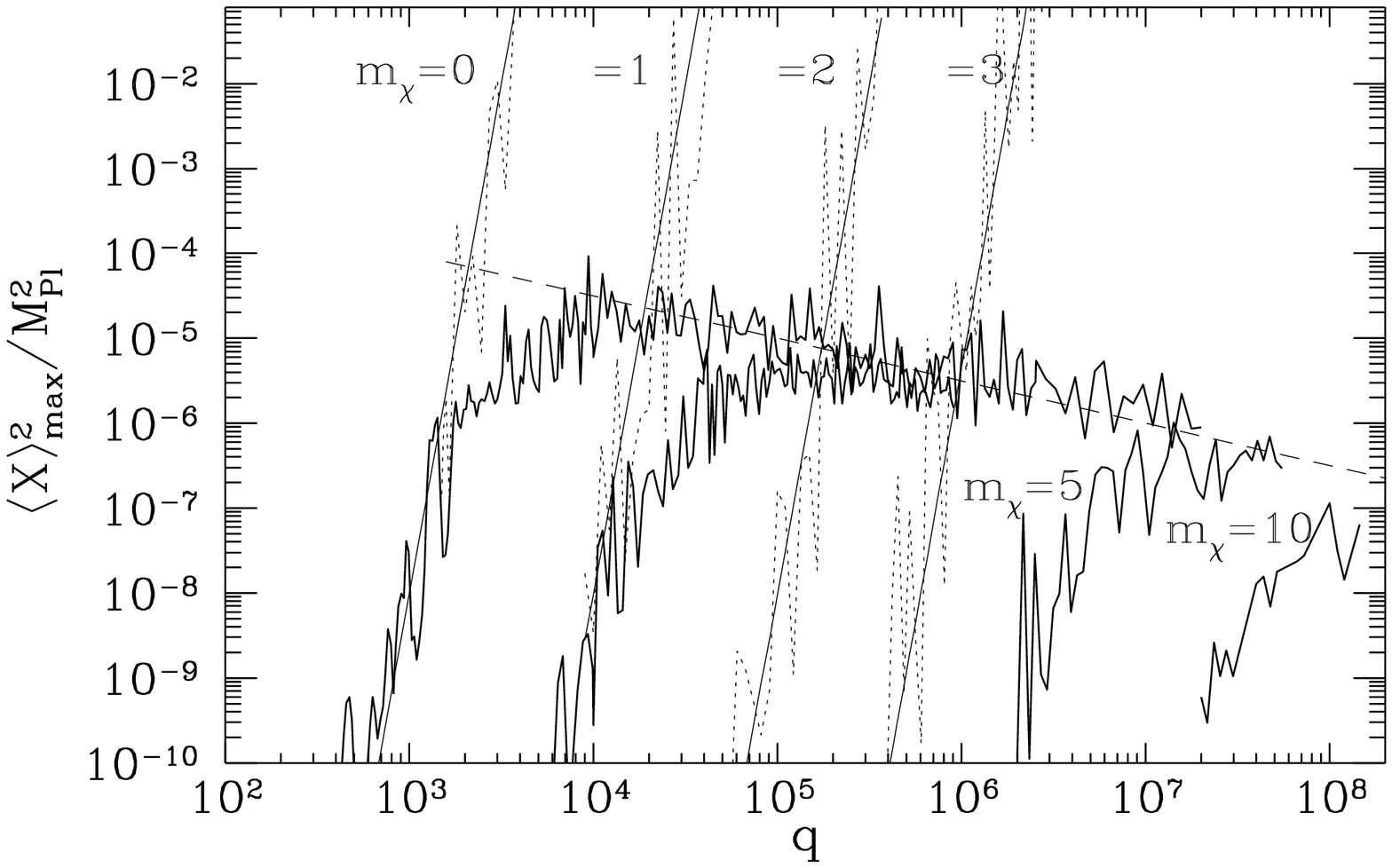,height=3.6in,width=5.6in}
\caption{Maximal value of the variance of fluctuations in Model 1
as a function of the resonance parameter. Dotted lines correspond to no
back reaction and are fitted by a power law. Solid curves are obtained
in the Hartree approximation. The dashed straight line is a 
power law fit to the Hartree data.}
\label{fig:Fig2}
\end{figure}

The resulting $\langle X^2 \rangle_{\rm max}$ as a function of $q$
(the brackets denote averaging over space) is shown in Fig. \ref{fig:Fig2}.
It is suppressed for large $q$ but is still not negligible. 
If we measure the strength of the (massive) fluctuations
with regard to symmetry restoration \cite{KLS2,Tkachev} using the effective 
``temperature": $T_{\rm eff}^2/12\equiv\langle X^2 \rangle_{\rm max}$, 
then for $q=10^8$, the Hartree approximation gives
$T_{\rm eff}$ of order $10^{-3} M_{\rm Pl}$.
This is of order of the GUT scale, so the process could be relevant for 
symmetry restoration even in Grand Unified models.

One should keep in mind, though, that the size of fluctuations can be
further altered (apart from the redshift) by the subsequent chaotic evolution
\cite{us}, absent in the Hartree approximation. 
That chaotic evolution is rather fast
because it is due to stimulated, i.e. classically enhanced, processes.
For example, in applications to GUT baryogenesis, the chaotic evolution
will in large part take place before the B-non-conserving decays of
$X$ particles. Therefore, some quantitative understanding of it is
required before reliable estimates of the baryon asymmetry can be made; this
will be a subject of a forthcoming paper.

Below we describe our models and calculations in more detail.

{\em Model 1: Massive inflaton}. In conformal variables, 
and after a convenient rescaling, the action for the matter fields is
\[
S_m = \frac{\phi^2(0)}{m^2} \int d^3 \xi d\tau
\left[
{1 \over 2} \left( \dot \varphi -\frac{\dot a}{a} \varphi \right)^2
                         - \frac{(\nabla \varphi)^2}{2}
+{1 \over 2} \left( \dot \chi -\frac{\dot a}{a} \chi \right)^2
                         -\frac{(\nabla \chi)^2}{2}
- {1 \over 2} \frac{a^2(\tau)}{a^2(0)} \varphi^2 
\right.
\]
\be
\left.
- {1 \over 2} \frac{M_X^2}{m^2} \frac{a^2(\tau)}{a^2(0)} \chi^2
- 2q \varphi^2 \chi^2
\right]   \,\, ,
\label{act1}
\ee
where $\tau=ma(0)\eta$, $\eta$ is the conformal time; $\xi=ma(0) x$;
$\varphi=\phi a(\tau)/\phi(0) a(0)$, 
$\chi=X a(\tau)/\phi(0) a(0)$; $a(\tau)$ is the scale factor of the Universe.

We take $\tau=0$ at the moment of the first extremum of $\phi(\tau)$.
Solving the equation of motion for the zero-momentum mode of the inflaton
coupled to gravity, together with the Einstein equation for the scale factor,
we have found $\phi(0)=0.28 M_{\rm Pl}$. At later times, we continued
to determine the evolution of the scale factor self-consistently, including
the influence of produced fluctuations in the Einstein equations. After 
$\tau=0$, a matter dominated stage quickly sets in, for which one can use
\be
\frac{a(\tau)}{a(0)}=\left(
\sqrt{\frac{\pi}{3}} \frac{\phi(0)}{M_{\rm Pl}} \tau + 1 \right)^2
\approx ( 0.29\tau + 1)^2 \,\, .
\label{sca1}
\ee
However, as our self-consistent solution confirms, for some values of the 
parameters the expansion later crosses over into a 
radiation dominated stage. That happens when the interaction potential
$\half g^2 \phi^2 X^2$ overtakes the mass term of $\phi$. 

In the Hartree approximation, 
the zero-momentum mode is the only non-vanishing mode of the field $\varphi$;
we keep for it the notation $\varphi$. Its equation of motion is
\be
{\ddot \varphi} - \frac{\ddot a}{a} \varphi 
+ \frac{a^2(\tau)}{a^2(0)} \varphi + 4q\langle \chi^2 \rangle \varphi = 0 \,\, ,
\label{eqm1}
\ee
where the brackets denote an average over space.
This should be solved with the initial conditions 
$\varphi(0)=1$, ${\dot \varphi}(0)=0$.
The equation of motion for the Fourier components of the field $\chi$ is
\be
{\ddot \chi}_{\k} + \omega^2_k(\tau) \chi_{\k} = 0 \,\, ,
\label{lin}
\ee
where
\be
\omega^2_k(\tau) = k^2 - \frac{\ddot a}{a} 
+\frac{M_X^2}{m^2} \frac{a^2(\tau)}{a^2(0)} 
+4q\varphi^2(\tau)  \,\, .
\label{ome}
\ee
We will sometimes use the notation
\be
m_{\chi}^2 \equiv \frac{M_X^2}{m^2} \,\, .
\label{mchi}
\ee
The Fourier components are normalized in such a way that the variance
of the fluctuation field is
\be
\langle \chi^2 \rangle = \frac{1}{(2\pi)^3}
\int d^3 k |\chi_{\k}|^2 \,\, .
\label{nor}
\ee
Because fluctuations are produced mostly with relatively small values of
(comoving) momenta $k$, the inevitable presence of a momentum cutoff in our
calculation, somewhere at larger momenta, will automatically result in
a properly renormalized $\langle \chi^2 \rangle$. If we wanted to embed
our calculation into a fully quantum-field-theoretical picture, we would
identify the parameters we use ($m$, $g^2$ etc.) with the running parameters 
of that picture, taken at a normalization point of the order of our momentum
cutoff; moderate changes in the normalization point will only cause 
corrections of relative order $g^2/4\pi = q m^2/4\pi\phi^2(0)$.

The initial conditions for $\chi_{\k}$ are randomly distributed with
the probability density
\be
{\cal P}_k[\chi_{\k}] \propto \exp\left[ -\frac{2\phi^2(0)}{m^2} \omega_k(0)
            |\chi_{\k}(0)|^2 \right] \,\,  .
\label{dis1}
\ee
According to eq. (\ref{act1}), $m^2/\phi^2(0)$ is the parameter that
regulates the strength of quantum fluctuations compared to
classical ones in this model.
The initial ``velocities" are locked to their corresponding ``coordinates"
\be
{\dot \chi}_{\k} (0) = [-i\omega_k(0) + h(0) ] \chi_{\k} \,\, ,
\label{vel}
\ee
where $h={\dot a}/a$.
For a derivation of these initial conditions, see ref. \cite{us}.
The semiclassical description is reliable as long as 
$g^2/4\pi=  q m^2/4\pi\phi^2(0)\ll 1$.

\begin{figure}
\psfig{file=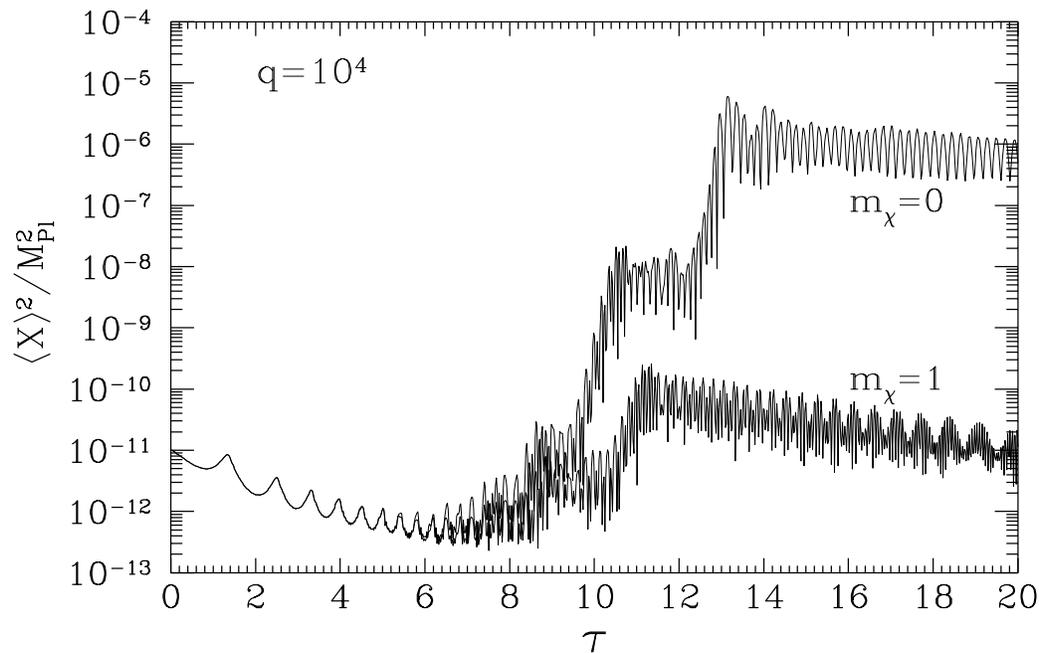,height=3.6in,width=5.6in}
\caption{Fluctuation variance in Model 1, as a function of time,
for two values of the mass of the fluctuating field; back reaction
of fluctuations on the inflaton field was included in the Hartree
approximation.}
\label{fig:Fig1}
\end{figure}

\begin{figure}
\psfig{file=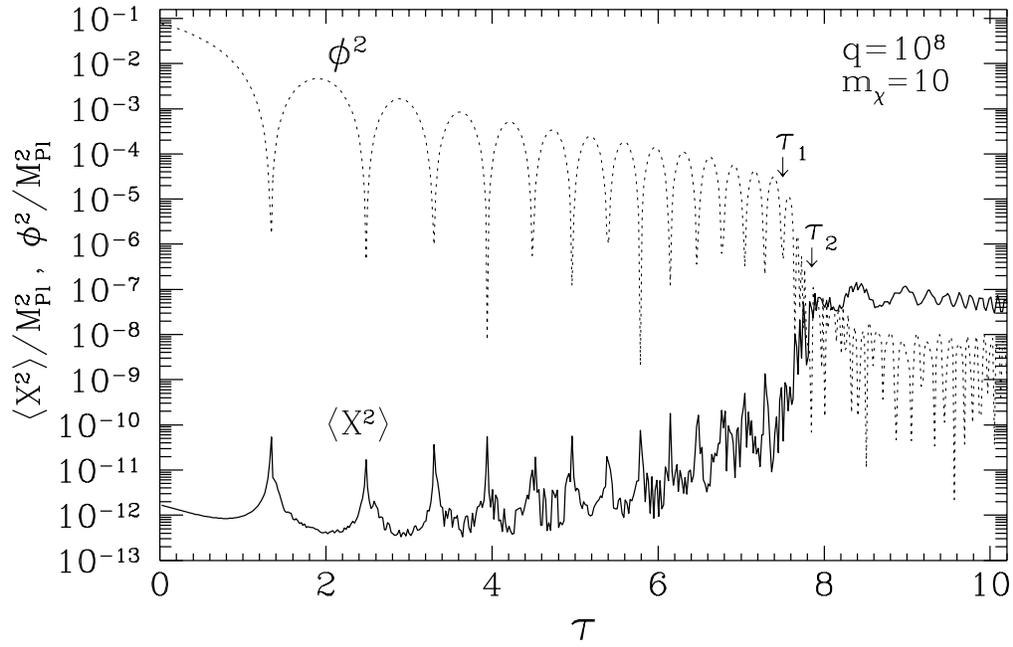,height=3.6in,width=5.6in}
\caption{Time evolution of the homogeneous inflaton field and the fluctuation
variance in Model 1 in the Hartree approximation, for given
values of the parameters.}
\label{fig:Fig3}
\end{figure}

The results of numerical calculations for the realistic case 
$m=10^{-6} M_{\rm Pl}$ are presented in Figs. \ref{fig:Fig2}--\ref{fig:Fig3}.

Three factors can potentially suppress a parametric resonance in the expanding 
Universe, each represented by the corresponding term in the equations of 
motion. In our conformal variables, they are:
the time dependent mass term $\propto M_X^2$
in eq. (\ref{ome}), which redshifts the modes away from the resonance; 
the time dependent mass term of the field
$\varphi$ in eq. (\ref{eqm1}), which makes the oscillations of 
the background field aperiodic; 
and the back reaction term $\propto q$ in eq. (\ref{eqm1}). 
A very useful quantity to introduce is the effective $q$,
\be
q_{\rm eff}(\tau) \equiv 
\frac{g^2 {\bar \phi}^2(\tau)}{4 m_{\rm eff}^2(\tau)} \,\, ,
\label{qeff}
\ee
where ${\bar \phi}$ is the (decreasing) amplitude 
of the physical field $\phi$,
and $m_{\rm eff}$ is that field's effective mass including the Hartree
correction. 
In a slowly expanding Universe (the adiabatic limit), $q_{\rm eff}(\tau)$
would indeed be the correct resonance parameter at time $\tau$.

To understand the effects of the redshift and the back reaction,
it is convenient to see first what happens when the back reaction is
neglected. This is indeed the correct limit for sufficiently small $q$
(a more precise criterion depends on $m_{\chi}$).
Then, $m_{\rm eff}=m$ and 
\be
q_{\rm eff}(\tau) = q\frac{{\bar \phi}^2(\tau)}{{\bar \phi}^2(0)}
\label{qeff1}
\ee
For a parametric resonance to be effective at time $\tau$,
$q_{\rm eff}(\tau)$ should not be much smaller than unity. 
Because a resonance will need some time $\tau_0$ to start, the condition for it 
to begin at all, $q_{\rm eff}(\tau_0)\gsim 1$, already requires large values 
of $q=q_{\rm eff}(0)$. By our definition,
a resonance starts when $\langle X^2 \rangle$ rises above quantum fluctuations. 
E.g. in Fig. \ref{fig:Fig1},
$\tau_0 \approx 6$; some $k$-modes can start to grow earlier,
however.

Consider first the case of massless $X$, $M_X=0$. We have found that
even then the resonance starts only for $q\gsim 10^3$, see Fig. 
\ref{fig:Fig2}. With $q \gsim 10^3$, the parametric
resonance starts at some point but terminates when $q_{\rm eff}$ drops
below unity. Hence, in the expanding Universe, the resonance stops
and the maximal size of produced fluctuations is limited in this
model\footnote{
As opposed to the conformally invariant case \cite{KLS,Kaiser,us}.}
even without back reaction.
We have calculated the maximum value of  $\langle X^2\rangle$ that
would be achieved without back reaction for a wide range of values of $q$.
It is plotted in Fig. \ref{fig:Fig2} by the dotted curve and for massless 
$X$ has the label $m_\chi =0$. This curve can be approximated by 
$\langle X^2\rangle_{\rm max} \propto q^{12}$, which 
is plotted in Fig. \ref{fig:Fig2} by a solid straight line. 

Next, consider the case $M_X\neq 0$. The rapidly growing mass term
of $\chi$ is a resonance suppressing factor.
In the case of the $\lambda
\phi^4$ model, we will see that it works against the resonance
extremely effectively.
In the present model, however, the frequency of oscillations
of the field $\varphi$ is growing at the same rate as the effective mass
of $\chi$, compare eqs.  (\ref{eqm1}) and  (\ref{ome}).
In a sense, massive $\chi$ modes can remain tuned to the resonance, in some
analogy with the conformal case. Still, for a non-zero $M_X$, the resonance 
terminates before $q_{\rm eff}$ drops below unity. 
This is why the curves with $m_\chi \ne 0$ are shifted in Fig. \ref{fig:Fig2}
to the right with respect to the $m_\chi = 0$ case. 

In the absence of the Universe expansion and back reaction, 
the linearized equation for fluctuations, eq. (\ref{lin}),
would be a Mathieu equation. 
In that case, a necessary condition for the $X$ particle 
production to fall into the strongest resonance band would be, for 
$m_{\chi}\gsim 1$: $m_{\chi}\lsim q^{1/4}$, or equivalently $q\gsim (M_X/m)^4$.
As one can see from Fig. \ref{fig:Fig2}, 
this estimate is four orders of magnitude off 
in the expanding Universe: e.g. with $q \sim 10^4$ the resonance barely 
develops even for $m_\chi=1$.
To take the Universe expansion into account, we can replace $q$ in the above
estimate with $q_{\rm eff}$ at some time before the end of the resonance stage, 
$q_{\rm eff}(\tau_1)$. As evident from eq. (\ref{qeff1}), finding 
$q_{\rm eff}(\tau_1)$ is equivalent to finding ${\bar \phi}(\tau_1)$.
The quantities $\tau_1$ and ${\bar \phi}(\tau_1)$ in general depend on $q$.
Although some analytical estimates of ${\bar \phi}(\tau_1)$ can be made
\cite{KLS2}, its reliable determination requires our numerical results.
We have found that for the range of $q$ we consider in Fig. \ref{fig:Fig2},
one can use generically 
${\bar \phi}^2(\tau_1)\sim 10^{-4} \phi^2(0)$.
Using $q_{\rm eff}$ instead of $q$ repairs the estimate:
to efficiently produce particles of mass $M_X$, one needs
$q_{\rm eff}(\tau_1)\gsim (M_X/m)^4$, or $q\gsim 10^4 (M_X/m)^4$, in agreement
with Fig. \ref{fig:Fig2}. For $M_X/m=10$, we get $q\gsim 10^8$, which for
$m=10^{13}$ GeV translates into $g^2\gsim 4.\times 10^{-3}$. Considering
the complexity of the phenomena involved, the estimate $g^2\gsim 10^{-4}$ 
made for this case in ref. \cite{KLR} is in a reasonable agreement with our 
result.

The term ``wide'' parametric resonance is to some extent misleading.
The resonance band itself is narrow, $\Delta k \ll q_{\rm eff}^{1/4}$.
However, the resonance band can be anywhere in the interval 
$0 < k_{\rm res} \lsim q_{\rm eff}^{1/4}$, 
and indeed moves through this interval
back and forth when $q_{\rm eff}$ slightly changes. Now, since $q_{\rm eff}$ 
is time dependent in expanding Universe, all $k$ from this range become
resonant in turn,\footnote{It is perhaps not obvious a priori that
this motion of the resonance band does not kill the resonance 
altogether.} and it is for that reason that one
may  consider $q_{\rm eff}^{1/4}$ as a kind of effective width of 
the resonance. As a result of the motion of the resonance band, 
all $k$ up to $q^{1/4}$ become filled up in the power spectrum, although to a 
varying extent.

In Fig. \ref{fig:Fig2}, notice also
the sensitivity of the maximal size of fluctuations with respect to small 
changes in $q$.
Small changes in $q$ can be simulated by small numerical
uncertainties, such as that of the coefficient in the relation
$\phi(0)=0.28 M_{\rm Pl}$. It anyway makes sense to
average out the rapid changes in Fig. \ref{fig:Fig2} 
and concentrate on the general pattern. As a general trend,
we observe a broad maximum of $\langle X^2 \rangle_{\rm max}$
as a function of $q$, at fixed $M_X$. It occurs in the region of $q$ where
the redshift and the back reaction play comparable roles in terminating
the resonance.

Now consider the limit when 
the maximum value attained by fluctuations is determined by the back reaction. 
This is the case of asymptotically large $q$.
The general trend in $\langle X^2
\rangle_{\rm max}$ in this region of $q$ is given by
\be
\langle X^2 \rangle_{\rm max} \sim \frac{3\times 10^{-3}}{\sqrt{q}}
M_{\rm Pl}^2
\label{num}
\ee
which is plotted in Fig. \ref{fig:Fig2} by the dashed straight line.
In this region, we can also estimate $\langle X^2 \rangle_{\rm max}$ 
in the Hartree approximation analytically, and the estimate is in rather
good agreement with eq. (\ref{num}). It proceeds as follows. We first
notice that because the decay of $\phi$ into $X$ is a stimulated process,
it mostly takes place in a short period close to the end of the resonance,
when $X$ fluctuations are already large. Then, we should 
distinguish between the amplitude of the inflaton oscillations just before
the decay, ${\bar \phi}(\tau_1)$, which is determined mostly by the redshift, 
and its amplitude right after the decay, ${\bar \phi}(\tau_2)$. 
For a wide resonance, ${\bar \phi}(\tau_2)$ can be smaller than
${\bar \phi}(\tau_1)$ by orders of magnitude. Such a drop is 
clearly seen in Fig. \ref{fig:Fig3} (even though for $m_{\chi}=10$ the value
$q=10^8$ is not an asymptotically large $q$).
Because $\tau_2-\tau_1$ is
such a short time, the Universe expansion during it can be neglected, 
and from energy conservation we get
\be
g^2 {\bar \phi}^2(\tau_2) \langle X^2\rangle 
\sim m^2 {\bar \phi}^2(\tau_1) \,\, .
\label{ene}
\ee
The second ingredient of our estimate is the observation \cite{KLS} that
the presence of $\langle X^2 \rangle$ changes the effective mass of the 
inflaton; the effective mass is given by 
$m^2_{\rm eff}=m^2+g^2 \overline{\langle X^2 \rangle}$, where 
$\overline{\langle X^2 \rangle}$ is a slowly changing part of
$\langle X^2 \rangle$. When $\langle X^2 \rangle$ becomes sufficiently 
large, 
$m^2_{\rm eff} \approx g^2 \overline{\langle X^2 \rangle}\sim 
g^2 \langle X^2 \rangle$, and the effective 
resonance parameter becomes $q_{\rm eff} = g^2 {\bar \phi}^2/4m_{\rm eff}^2 
\sim {\bar \phi}^2/\langle X^2 \rangle$. 
The resonance stops when $q_{\rm eff}(\tau_2)\sim 1$; using eq. (\ref{ene}),
this gives\footnote{
Our usual condition $q_{\rm eff}\gsim (M_X/m_{\rm eff})^4$ is satisfied
for $q_{\rm eff}\sim 1$ at sufficiently large $q$, due to the
large $m_{\rm eff}$.
Notice also that because a large $m_{\rm eff}$ lowers the $M_X/m_{\rm eff}$ 
ratio, the Hartree correction to the mass of $\phi$
actually helps the resonance for a while, as we 
have indeed observed in our numerical results.}
\be
\langle X^2 \rangle_{\rm max} \sim \frac{m{\bar \phi}(\tau_1)}{g}
\label{ana}
\ee
The value of ${\bar \phi}(\tau_1)$ is weakly depends on $q$.
As before, we use the generic estimate 
${\bar \phi}(\tau_1)\sim 10^{-2} \phi(0)$ for all $q$ considered.
We then obtain 
$\langle X^2 \rangle_{\rm max} \sim 10^{-3} M_{\rm Pl}^2/\sqrt{q}$,
in reasonable agreement with the numerical result (\ref{num}).
A $1/\sqrt{q}$ (up to logarithms) estimate for $\langle X^2 \rangle_{\rm max}$ 
appears in refs.\cite{KLS,KLS2,ALR}, although the way the estimate is
obtained and the numerical coefficient are somewhat different from ours. 
We stress again that we do not know at present if the $1/\sqrt{q}$ behavior 
is an artifact of the Hartree approximation in this model. 
See also the discussion of the conformally invariant case below.

Note that at asymptotically large $q$, 
the interaction term $\half g^2 X^2 \phi^2$
dominates over the mass term of $X$ during the entire
resonance stage. So, at the resonance stage, the fluctuations of $X$ 
cannot really be described as 
particles, since they are strongly coupled to the oscillating background. 

{\em Model 2: Massless inflaton}. In rescaled variables, the action for 
the matter fields in this model is
\[
S_m=\frac{1}{\lambda} \int d^3 \xi d\tau
\left[
{1 \over 2} \left( \dot \varphi -\frac{\dot a}{a} \varphi \right)^2
                         - \frac{(\nabla \varphi)^2}{2} 
+{1 \over 2} \left( \dot \chi -\frac{\dot a}{a} \chi \right)^2
                         -\frac{(\nabla \chi)^2}{2}
- {1 \over 4} \varphi^4
\right.
\]
\be
\left.
- {1 \over 2} \frac{M_X^2}{\lambda \phi^2(0)} \frac{a^2(\tau)}{a^2(0)} \chi^2
- 2q \varphi^2 \chi^2
\right]
\label{act2}
\ee
where $\varphi$ and $\chi$ are defined as before but
$\tau=\sqrt{\lambda} \phi(0) a(0) \eta$, 
$\xi=\sqrt{\lambda} \phi(0) a(0) x$.

We again take $\tau=0$ to be the first extremum of $\phi(\tau)$.
In this case we find $\phi(0)=0.35 M_{\rm Pl}$.
There is no resonance for a massive enough $X$ (see below).
In that case, after the inflaton oscillations start,
the Universe quickly becomes radiation dominated and we can use
\be
\frac{a(\tau)}{a(0)}=
\sqrt{\frac{2\pi}{3}} \frac{\phi(0)}{M_{\rm Pl}} \tau + 1 
\approx 0.51\tau + 1
\label{sca2}
\ee
All other formulas are obtained from the corresponding ones for Model 1
by replacing $m^2$ with $\lambda\phi^2(0)$.
Accordingly, we now use $m_\chi=M_X/\sqrt{\lambda}\phi(0)$.

For a massless field $\chi$ the problem becomes conformally equivalent to 
the one in the Minkowski space-time: 
the expansion of the Universe decouples in the conformal coordinates. 
The parametric resonance is always effective. We have done a complete 
non-linear analysis of this problem
in the range $1 \le q \le 1000$, using the approach developed in 
ref. \cite{us}. At  $ q \gg 1 $, the evolution of the system is fast and we 
were able to trace the semiclassical thermalization
\cite{us} up to a slowly evolving state with a developed Kolmogorov 
turbulence. These results will be described in more detail
elsewhere; for previous discussions of kinetic phenomena after
preheating, see refs. \cite{us,Son2}. 
For the limited set of values of $q$ for which we now have the data,
$\langle X^2 \rangle$ at the end of the resonance stage in the 
full problem fits a $1/q$ curve. If confirmed by more extensive 
calculations, such behavior would be in disagreement with the approximately 
$1/\sqrt{q}$ dependence obtained in the Hartree approximation. 

With a massive $\chi$ field we have not found any parametric resonance 
for $m_\chi > 1$ and $q$ up to $q =10^6$. Analytically, the condition for
the resonance to exist at time $\tau$ (neglecting back reaction)
is $q^{1/4}\gsim m_{\chi} a(\tau)/a(0)$.
In the conformally invariant case ($m_{\chi}=0$), the fluctuations reach
appreciable strength at
$\tau\sim 100$. Using that value in the estimate for $m_{\chi}\neq 0$, 
we would expect that for $m_\chi=1$, the resonance exists when $q\gsim 10^{7}$.
We indeed observed a slow (physically insignificant) resonance at
$m_\chi =1$ and $q=10^8$.

In conclusion, we have studied the decay of massive and massless inflatons into
massive excitations in the expanding Universe. 
We have found that a wide
enough resonance can survive in the expanding Universe, though the expansion
is very important for determining precisely how wide it should be.
For a massive inflaton, the effective production of particles with mass ten
times that of the inflaton requires very large values of the resonance
parameter $q$, $q\gsim 10^8$. For these large $q$, the maximal size of produced
fluctuations is significantly suppressed by the back reaction, but at least
within the Hartree approximation they are still not negligible.
We do not know at present if this conclusion will persist in the fully
non-linear picture; the full account for rescattering is thus an important
problem for the future. For the massless inflaton with a $\lambda\phi^4/4$
potential, the Universe expansion completely prevents a resonance production
of particles with masses larger than $\sqrt{\lambda}\phi(0)$
for $q$ up to $q=10^6$.

We thank G. Dvali, L. Kofman, E. Kolb, A. Linde, A. Riotto, V. Rubakov, 
M. Shaposhnikov, and N. Tetradis for discussions and comments. S.K. thanks
CERN Theory Division, where some of the discussions took place, for
hospitality. The work of S.K. was supported
in part by the U.S. Department of Energy under Grant DE-FG02-91ER40681 
(Task B), by the National Science Foundation under Grant PHY 95-01458, 
and by the Alfred P. Sloan Foundation. The work of I.T. was supported
by DOE Grant DE-AC02-76ER01545 at Ohio State.

\def\baselinestretch{.7}

\end{document}